\documentclass[%
 reprint,
 amsmath,amssymb,
 aps,
floatfix,
]{revtex4-2}

\usepackage{graphicx}
\usepackage{dcolumn}
\usepackage{bm}
\usepackage{color}
\usepackage{float}
\usepackage{subcaption}
\usepackage{multirow}
\usepackage[all]{nowidow}

\newcommand{\unige}{Universit\'{e} de Gen\`{e}ve, Section de Physique, DPNC, 1205 Gen\`{e}ve, Switzerland}

\begin{document}

\title{ The Superscaling Variable and Neutrino Energy Reconstruction,\\ From Theoretical Predictions to Experimental Limitations}

\author{D.~Douqa}
\author{S.~Bordoni}
\author{L.~Giannessi}
\affiliation{\unige}
\author{A.~Nikolakopoulos}
\affiliation{Theoretical Physics Department, Fermilab, Batavia, IL 60510, USA}
\author{F.~S\'anchez}
\author{C.~Schloesser}
\affiliation{\unige}

\date{\today}
             
\begin{abstract}
We introduce the novel approach of using the superscaling variable as an observable and an analysis tool in the context of charged current neutrino-nucleus interactions. We study the relation between the superscaling variable and the removal energy, in addition to other fundamental parameters of the neutrino-nucleus interaction models. In the second half of the paper, we discuss the experimental viability of this measurement following a study of neutrino energy and missing momentum reconstruction. We show that the superscaling variable is measurable in neutrino interaction experiments provided that the proton is detected in the final state. We discuss the resolution of this measurement, and the limitation imposed by the proton's detection threshold.
\end{abstract}

\maketitle

\section{Introduction}
The studies of neutrino-nucleus interactions are entering a new stage, motivated by long-baseline experimental programs, in which the statistical uncertainties will diminish and thus nuclear effects -- contributing to the systematic error -- have to be kept well under control~\cite{Alvarez-Ruso:2017oui}. The incomplete theoretical knowledge of neutrino-nucleus interactions influences various stages of experimental analyses. For instance, for the future Hyper-Kamiokande water Cherenkov detector~\cite{Hyper-Kamiokande:2018ofw}, the method for reconstructing the neutrino energy will be mainly based on the kinematics of the outgoing muon, which is the only particle observed, assuming that the reaction-mechanism is two-body quasi-elastic (QE) scattering on a bound nucleon.

In the context of CCQE interactions, the superscaling approach has been advocated to enhance some of the most advanced models of neutrino-nucleus interactions~\cite{Gonzalez-Jimenez:2014eqa}. 
Superscaling was introduced initially to describe electron-nucleon scattering where it is a very well established feature observed in the data~\cite{Donnelly:1999sw,Barbaro:1998gu}. The experimental determination of the superscaling behaviour of neutrino interactions will be fundamental to the development of the aforementioned models. 

There are many uncertainties in the modelling of neutrino-nucleus interactions~\cite{Alvarez-Ruso:2017oui}, among them the description of the initial state nucleus, dynamical phenomena such as Random Phase Approximations (RPA) or 2p2h interactions, Pauli blocking and nuclear re-scattering. The removal energy —energy absorbed by the unobserved nuclear system during the interaction— has emerged among them as one of the largest systematic errors on T2K neutrino oscillations~\cite{ systUncertainties}. A direct measurement of the average removal energy will be very beneficial to reduce systematic errors and to add credibility to neutrino-nucleus interaction models. However, in flux-averaged data, the removal energy is not directly observable as the neutrino energy is unknown on an event-to-event basis. We present the reconstructed superscaling variable as a potential observable to address this uncertainty experimentally.

We introduce theoretical predictions and discuss the relation between the superscaling variable and the removal energy, employing a range of neutrino-nucleus interaction models. These models span from the conventional Relativistic Fermi Gas (RFG), for which the superscaling variable was originally formulated, to more sophisticated Mean Field approaches. We perform a study under the conditions of the T2K experiment~\cite{T2KExp}, using the so-called topological event selection. Topologies are based on the counting of particles emerging from the nucleus after nuclear re-scattering (NrS). In this study we concentrate on the CC0$\pi$1p topology, with a muon, a proton and no mesons in the final state. We show that a measurement of the reconstructed superscaling variable provides sensitivity to the removal energy in a $1$-nucleon knockout event (CC1p1h).

\section{Event Kinematics and Neutrino Energy Reconstruction}
\label{sec:kinem}
In this section we discuss the kinematics of the CC1p1h reaction, and the neutrino energy reconstruction in a $1\mu1p$ event. We use several nuclear models for this study, and expose the differences between their treatments.

We consider the semi-inclusive reaction, in which a proton and a muon are detected in coincidence
\begin{equation}
\nu_\mu(k_\nu) + A(k_A) \rightarrow \mu(k_\mu) + p(k_p) + B(k_B)
\end{equation}
where $A$ is the initial nucleus, and $B$ is the residual hadronic system, which remains undetected. Particle four-vectors are labelled between parentheses, where $k_i = (E_i = \sqrt{M_i^2 + \lvert\vec{k}_i\rvert^2}, \, \vec{k}_i)$.

The energy and momentum transfer to the nucleus are $\omega = E_\nu - E_\mu$, and $\vec{q} = \vec{k}_\nu - \vec{k}_\mu$, respectively. 

Momentum and energy conservations are expressed as
\begin{eqnarray}
    \vec{q} &=& \vec{k}_p + \vec{k}_B \\
    \omega - E_p &=& M_B - M_A + \xi^* + T_B
\end{eqnarray}
where $\xi^*$ is the excitation energy of the final nucleus, $M_B$ is its ground state mass and $T_B$ is its kinetic energy. A further simplification can be performed using the definition of the experimental nucleon separation energy $S_N = M_B - M_A + M_N$, giving
\begin{equation}
\omega - E_p + M_N = S_N + \xi^* + T_B
\end{equation}

The missing momentum is defined as $\vec{p}_m \equiv -\vec{k}_B$, while the missing energy is defined as
\begin{equation}
E_m = \omega - T_p - T_B= S_N + \xi^* + (M_p - M_N)
\end{equation}

For what follows, it is useful to define 
\begin{eqnarray}
\tilde{E}_m &=& E_m - (M_p - M_N) \\
\label{eq:removalEn}
 &=& \omega - E_p + M_N \\
 \label{eq:S_def}
 &\equiv& S_N + \xi^*
\end{eqnarray}
where we neglect the kinetic energy of the recoiling nucleus as it becomes small when the mass of the residual system is large with respect to its momentum. The missing energy can be associated with the removal energy $S$, the energy needed to remove a nucleon from the nucleus, and we will treat them as equivalent ($S \equiv \tilde{E}_m$). 

\subsection{ Monte Carlo Models }
\label{sec:models}
NEUT~\cite{Hayato_2021} provides a very good environment to predict experimental observables. It includes a complete simulation of resonance production, shallow and deep inelastic scattering, meson exchange currents and charged and neutral currents both for neutrinos and anti-neutrinos, and for several nuclei. In addition, it includes an intra-nuclear cascade model, developed in \cite{Salcedo:1987md} and recently tuned to available scattering data \cite{PinzonGuerra:2018rju}.

In order to study the model-dependence of our procedure, we use the following interaction models implemented in NEUT for QE scattering. 
\begin{enumerate}
    \item The Relativistic Fermi Gas (RFG) model, which uses the Smith-Moniz parametrization~\cite{SMITH1972}.
    
    \item The Local Fermi Gas (LFG) model of Refs.~\cite{Nieves:2004wx, Gran:2013kda}. It includes corrections for long-range correlations, final-state interactions, and the Coulomb potential experienced by the charged lepton. The final-state nucleon kinematics are generated from the LFG momentum distribution assuming quasi-free scattering, as described and applied to neutrino scattering data in Ref.~\cite{Bourguille:2020bvw}.   
    
    \item The Spectral Function (SF) model. This is a calculation in the Plane Wave Impulse Approximation (PWIA) that uses the realistic Rome spectral function~\cite{BENHAR1994493, PhysRevD.72.053005}. The implementation is detailed in Ref.~\cite{Furmanski:2016wqo}.
\end{enumerate}

In addition to the NEUT native models, we include the Energy-Dependent Relativistic Mean Field (EDRMF)~\cite{Gonz_lez_Jim_nez_2019} and Relativistic Plane Wave Impulse Approximation (RPWIA)~\cite{Gonzalez-Jimenez:2021ohu} models where NEUT's intra-nuclear cascade was used to simulate nuclear re-scattering and final state interactions (FSI) following the same approach presented in \cite{Nikolakopoulos:2022qkq}.

Within the Fermi gas models, interaction with an initial on-shell nucleon with momentum $\vec{k}_N$ is considered. In this case $\vec{p}_m = \vec{k}_N$ and
\begin{equation}
\label{eq:RFGconsv}
\omega - E_p + M_N = T_F - T_N + S_N
\end{equation}
where $T_F$ is the kinetic energy of a nucleon at the Fermi level. One thus has
\begin{equation}
\label{eq:RFGpmEm}
\tilde{E}_m = T_F - T_N + S_N
\end{equation}
where for the RFG model, $T_F$ is a constant. 

On the other hand, in the LFG model, $T_F$ depends on the local nuclear density, and an average over the density is performed. The value of $T_F$ is hence smeared out, losing the direct relation between missing energy and missing momentum of the RFG.

The RPWIA and EDRMF approaches~\cite{Nikolakopoulos:2022qkq, Gonzalez-Jimenez:2021ohu, Franco-Patino:2022tvv} use a RMF model to describe the initial state~\cite{NLSH, Horowitz81}.
In this case, a nucleon is knocked out from a nuclear shell, which has a fixed missing energy.
For carbon, considered in this work, these are the $s_{1/2}$ and $p_{3/2}$ shells, with occupancies of two and four nucleons respectively.

The SF calculation uses a more realistic energy and momentum density, by implementing the experimental observation that the energy of the nuclear shells is smeared out, and that the shell-model states are depopulated~\cite{BENHAR:RevModPhyseep,RDWIA:Kelly, Dutta:PRCeepAu}.
The strength missing from the mean-field appears at high missing energy and momentum and is modelled in the local-density approximation~\cite{BENHAR1994493, PhysRevD.72.053005}.

These approaches represent an increasingly realistic description of the nuclear spectral function. And the EDRMF results, that use the same RMF initial state as the RPWIA, are used to gauge the effect of nucleon final state interactions. Consistent comparisons in the PWIA using these different spectral functions, and detailed discussion on the kinematics, are presented in Ref.~\cite{VanOrdenDonnelly:PWIACC}.

\subsection{Reconstruction of Full Kinematics}
\label{sec:recon}
The full kinematics of the CC1p1h event, where the target nucleon is a neutron ($\nu_\mu + n \rightarrow \mu + p$), can be reconstructed using two different approaches. When the missing energy and momentum are fully correlated, as in the RFG model, we can perform reconstruction by energy-momentum conservation (REP). In this case, Eq.~\ref{eq:RFGconsv} can be solved for the longitudinal missing momentum, giving
\begin{equation} 
     p_m^L = \frac{ (p_{m}^T)^2+ M_N^2 - (E_{\mu} + E_{p} - k_{\mu}^L - k_{p}^L + S_{\rm REP} )^2}{2 (E_{\mu} + E_{p} - k_{\mu}^L - k_{p}^L + S_{\rm REP} ) },
\end{equation}
where the superscript $L$ denotes the components in the direction of the neutrino beam, and
\begin{equation}
     p_m^T = (\vec{k}_{p} + \vec{k}_{\mu})^T,
\end{equation} 
is the missing momentum transverse to the beam. The neutrino energy can then be obtained, using the fact that $E_\nu = |\vec{k}_\nu|$, from
\begin{equation} 
     E_{\nu} = k_{\mu}^L + k_{p}^L - p_{m}^L
\end{equation}

Here, $S_{\rm REP}$ is a constant representing the average $\overline{\omega - E_p + E_N}$ (or $\overline{T_F + S_N}$), which in the RFG model can be identified as the fixed removal energy.

\begin{figure}[h] 
\begin{center}
\includegraphics[width=0.45\textwidth]{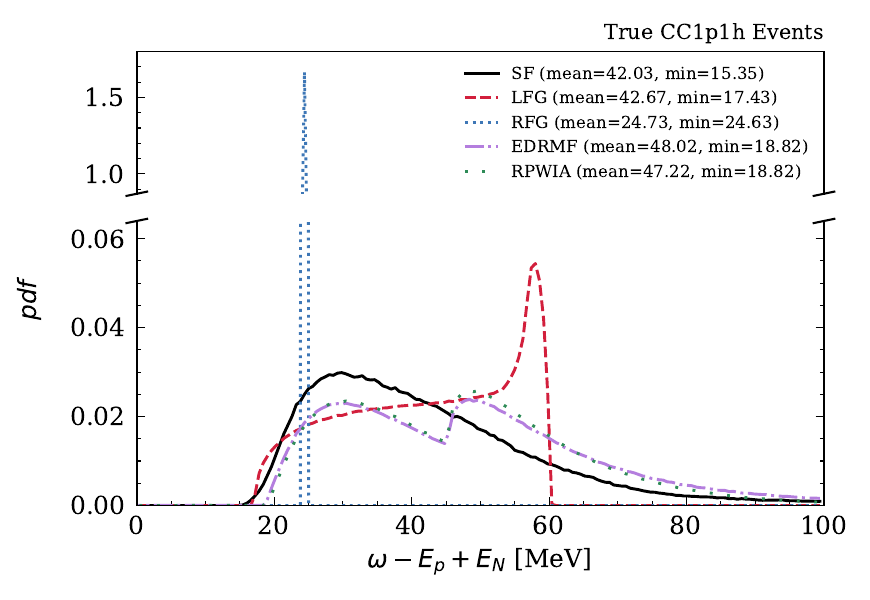}
\includegraphics[width=0.45\textwidth]{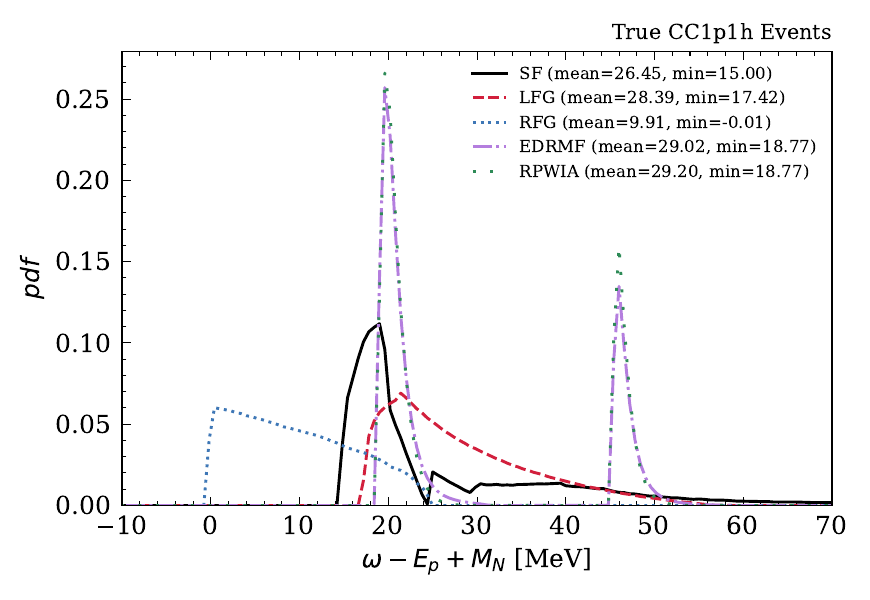}
\caption{The true distributions of $\omega - E_p + E_N$ (REP) and $\omega - E_p + M_N$ (RE) for the five models for true CC1p1h events.}
\label{Fig:TFTN}
\end{center}
\end{figure}

The distribution of $S_{\rm REP}$ is shown in the top panel of Fig.~\ref{Fig:TFTN}. It is clear that while for the RFG model this leads to an exact reconstruction, this is not the case for the other more realistic models that we consider.

A more realistic approach, is to consider $\tilde{E}_m \approx constant$, hence performing reconstruction solely by energy conservation (RE)
\begin{equation}
    E_{\nu} = E_{\mu}+ E_{p} + S_{\rm RE} - M_N
\label{eq:ene-recon}
\end{equation}

with $S_{\rm RE}$ being a constant, which corresponds to the average missing (removal) energy $\overline{\omega - E_p + M_N}$ (or $\overline{S_N + \xi^*}$).

This assumption considers $\tilde{E}_m$ to be independent from the missing momentum, which is natural for the RPWIA, EDRMF and SF approaches in the low missing (removal) energy region, where the bound-states do not satisfy a dispersion relation based on free nucleon kinematics. The only dependence of $\tilde{E}_m$ on the missing momentum comes from the recoil energy of the residual nucleus. This is seen clearly in the bottom panel of Fig.~\ref{Fig:TFTN}, where the tails of the peaks in the EDRMF and RPWIA calculations are due to the recoil energy. In the LFG model, the average $\tilde{E}_m$ corresponds to $\overline{T_F-T_N+S_N}$ from Eq.~\ref{eq:RFGpmEm}. Since the value of $T_N$ in the LFG model peaks around the local value of $T_F$, there is a smaller dispersion of the average $\overline{T_F-T_N+S_N}$ compared to $\overline{T_F+S_N}$, which can be observed in Fig.~\ref{Fig:TFTN}. 

This implies that the RE method would perform better in the reconstruction of the neutrino energy for the SF, LFG, EDRMF and RPWIA models compared to the REP method, the latter being only suitable for the RFG.

The average removal energy is computed from the averages of the distributions in Fig.~\ref{Fig:TFTN}. Taking into consideration the optimal reconstruction method for each model, the SF, LFG, EDRMF and RPWIA models have $S_{\rm RE}$ values of 26.45, 28.39, 29.02, and 29.20~MeV, respectively. On the other hand, the RFG model has an $S_{\rm REP}$ value of 24.73~MeV. 

To perform the neutrino energy reconstruction, we need to make an assumption for a constant average removal energy. Based on the distributions in Fig.~\ref{Fig:TFTN}, we set $S_{\rm RE} = S_{\rm REP} = 28$~MeV for all models. We will depart from this assumption later in the paper to explore the impact of changing the average removal energy on the superscaling variable distribution.

\begin{figure}[h]
\begin{center}
\includegraphics[width=0.45\textwidth]{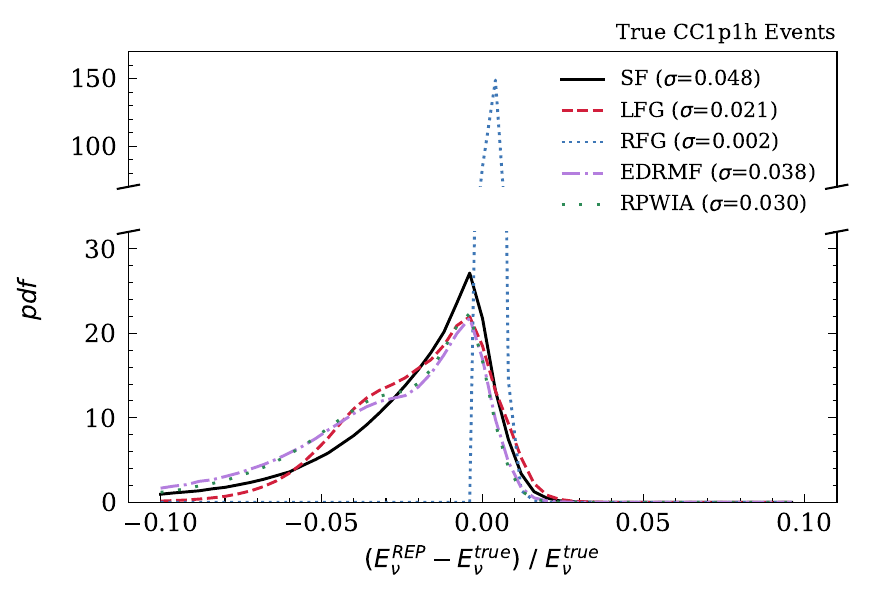}
\includegraphics[width=0.45\textwidth]{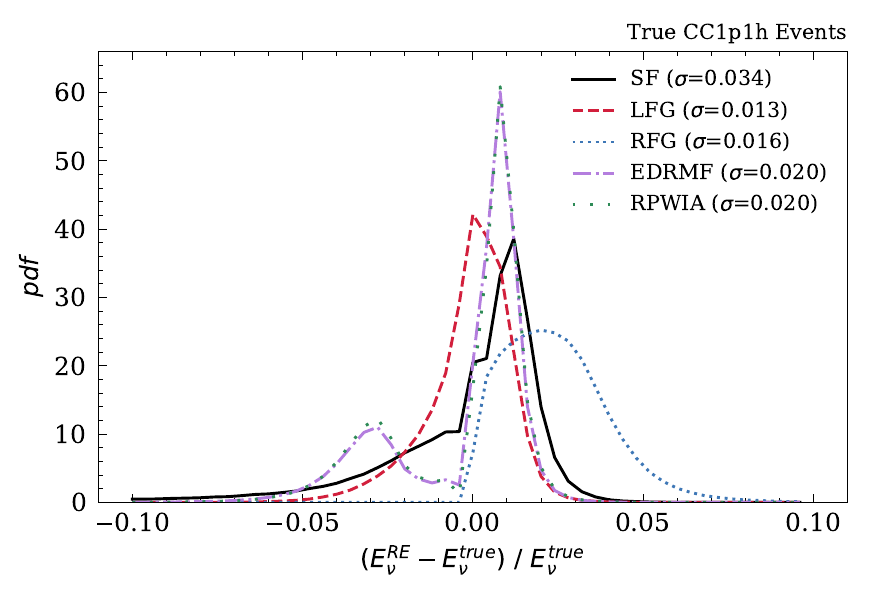}
\caption{The reconstructed neutrino energy resolution via REP (top) and RE (bottom) for true CC1p1h events.}
\label{Fig:ErecoComp}
\end{center}
\end{figure}

The results for the reconstructed neutrino energy, using the RE and REP methods are shown in Fig.~\ref{Fig:ErecoComp}. For the RE method (bottom panel), the distributions are mainly contained between $-4\%$ and $2\%$ for the LFG and SF models, while the EDRMF and RPWIA have a distinct concentration of strength at $-3\%$ due to the $s$-shell contribution. The REP method, shown in the top panel, exhibits a much broader but very similar distribution for all these models. However, the RFG is the outlier, with a significantly better resolution in REP, as expected.

\begin{figure}[h]
\begin{center}
\includegraphics[width=0.45\textwidth]{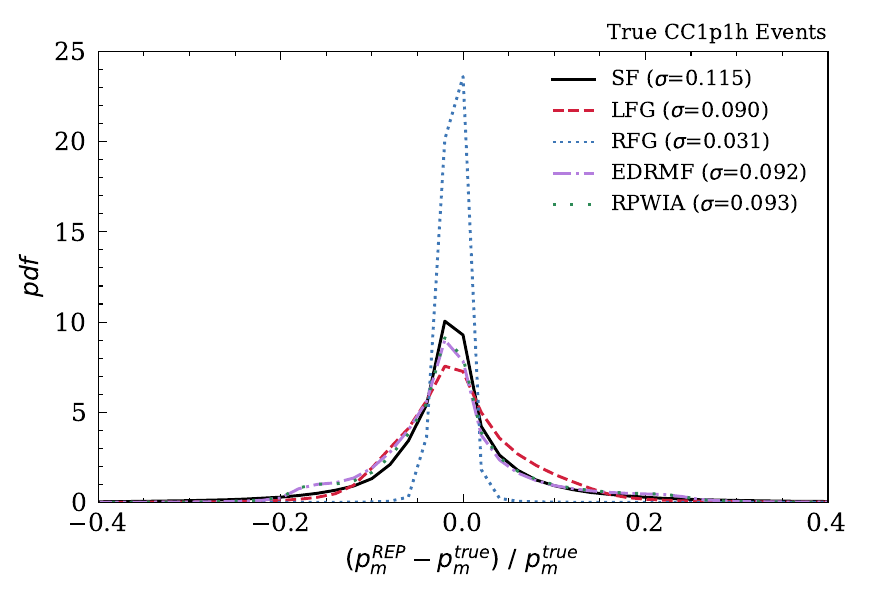}
\includegraphics[width=0.45\textwidth]{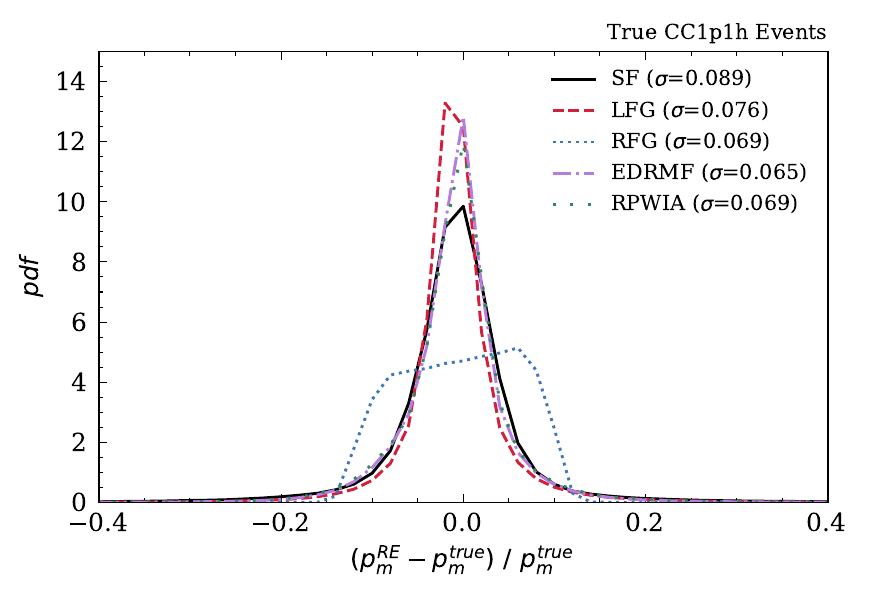}
\caption{The reconstructed missing momentum resolution via REP (top) and RE (bottom).}
\label{Fig:PnrecoComp}
\end{center}
\end{figure}

The reconstructed missing momentum is shown in Fig.~\ref{Fig:PnrecoComp}, again the RFG is the outlier, while for the other models we obtain fairly symmetric distributions for RE, with a full width at half maximum of around $5\%$.

It is clear that the RE method is more robust and provides better results for most models. Therefore, we will be using the RE method for all models, keeping in mind that the RFG model is a special case due to its simplicity.

Lastly, we note that these assumption will become less reliable when the signal has a large contribution from processes other than one-nucleon knockout. However, as we will show in section~\ref{sec:expEvSel}, the bulk of these contributions can be removed by a kinematic cut on the reconstructed missing momentum.

\section{ The Superscaling variable }
When an interaction between a particle and a many-body system involves energy and momentum transfers only to individual constituents of the complex system, the inclusive cross section can be approximated as a single-nucleon cross section times a specific function of the energy and momentum transfers $f(\omega, |\vec{q}|)$. Scaling, of the first kind, occurs when that function becomes independent of both $\omega$ and $\vec{q}$ explicitly. The scaling function depends on the kinematics only through a single quantity $\psi(\omega, |\vec{q}|)$~\cite{doi:10.1146,Amaro_2020}. 

The superscaling variable $\psi'$ was first introduced by Donnelly and Sick \cite{Donnelly:1999sw} as an evolution of the scaling variable $\psi$ used by Alberico et al. \cite{alberico:1988} and within the framework of the RFG model. It is defined as: 

\begin{equation}
    \psi'(\omega, \vec{q}) =  \frac{1}{\sqrt{\sqrt{1+\eta_F^2 } - 1}} \frac{\lambda - \tau} {\sqrt{(1+\lambda)\tau + \kappa \sqrt{\tau(1+\tau) }} }   
    \label{eq:psi}
\end{equation} 
with 
\begin{eqnarray}
    \label{eta_def}
     \eta_F &=& \frac{k_F}{M_N }  \\
     \label{kappa_def}
     \kappa &=& \frac{|\vec{q}\,|}{2 M_N} \\
     \label{lambda_def}
 \lambda &=& \frac{\omega - E_{\rm \rm shift}}{2 M_N } \\
 \label{tau_def}
 \tau &=& \kappa^2-\lambda^2 
\end{eqnarray}
where, $M_N$ is the neutron mass, $k_F$ is the Fermi momentum (fixed to 228~MeV for Carbon~\cite{PsiParameters}) and $E_{\rm shift}$ is a shift energy used to make the quasi-elastic peak coincide with $\psi' = 0$. Presumably, $E_{\rm \rm shift}$ includes information on the separation energy, the average removal energy of nucleons in the nucleus as well as aspects of final-state interactions like RPA, which can influence the removal energy distribution by altering the cross-section strength as a function of the momentum transfer in the reaction. In the following, we use the reported value for Carbon of 20~MeV~\cite{PsiParameters}.

The discussion in the following sections relies on the analysis of the special case when $\psi'=0$. This condition is achieved when: 
\begin{eqnarray}
    \lambda - \tau = \lambda - (\kappa^2 - \lambda ^2 ) = 0 
\label{Eq"psi0condition}
\end{eqnarray}

The $\lambda$ parameter has a direct dependency with $E_{\rm shift}$ and the removal energy $S$—assumed to be equivalent to the missing energy $\tilde{E}_m$ in section~\ref{sec:kinem}—through the energy transfer $\omega$. From Eq.~\ref{eq:removalEn}:
\begin{eqnarray}
    \omega = E_{\nu}-E_{\mu} = E_p + S - M_N 
\end{eqnarray}

By solving Eq.~\ref{Eq"psi0condition} and considering the positive solution as the physically meaningful one, we arrive at the following relationship
\begin{eqnarray}
    S - E_{\rm shift} = \sqrt{ M_N^2 + |\vec q|^2} - E_p  
\label{Eq:ZeroCondition}
\end{eqnarray}

For vanishing missing momentum, Eq.~\ref{Eq:ZeroCondition} becomes 
\begin{equation} 
\begin{split}
    S - E_{\rm shift} &= \sqrt{ M_N^2 + |\vec k_p|^2 } - E_p \\
     & \approx \frac{1}{2} \frac{M_N^2-M_p^2}{E_p} 
\end{split}
\label{eq:S-Eshift_noFermi}
\end{equation}
where it is clear that $E_{\rm shift}$ approaches the average value of $S$ (previously denoted as $S_{\rm RE}$) for $|\vec{p}_m| = 0$, with a maximum difference of 1.29~MeV, attained when the proton is at rest.

On the other hand, for $|\vec{p}_m| > 0$, one obtains
\begin{eqnarray}
    S - E_{\rm shift} \approx \frac{p_m^2 - 2\vec{k}_p\cdot\vec{p}_m}{2E_p}
\label{Eq:SN-Es}
\end{eqnarray}
where the equation remains more complex with dependencies on the missing and final nucleon momenta through the residual dependency on $|\vec q|$.

As such, $E_{\rm shift}$ should be a good estimator, within a few MeV, of the average removal energy, at low missing momenta.

Note that in the analysis of inclusive electron scattering data~\cite{PsiParameters}, the value of $E_{\rm shift}$ is determined by requiring that $\psi^\prime = 0$ at the experimentally observed quasi-elastic peak. In the RFG, the peak of the cross section is indeed obtained for $\psi' = 0$ when $E_{\rm shift}$ is equal to the separation energy used in the model. The presence of final-state interactions, RPA effects, and Pauli blocking among others, can break this simple relation. In any case, the peak position of $\psi^\prime$ is still strongly correlated with the average removal energy. In the following section, we study in detail the distribution of the superscaling variable, and its dependence on kinematic variables.

\section{True Monte Carlo Predictions}
\label{sec:truePred}
In this section, we perform a $\psi'$ calculation and characterization using samples of true CC1p1h events, without including nuclear re-scattering (NrS). 

\begin{figure}[h]
\begin{center}
\includegraphics[width=0.45\textwidth]{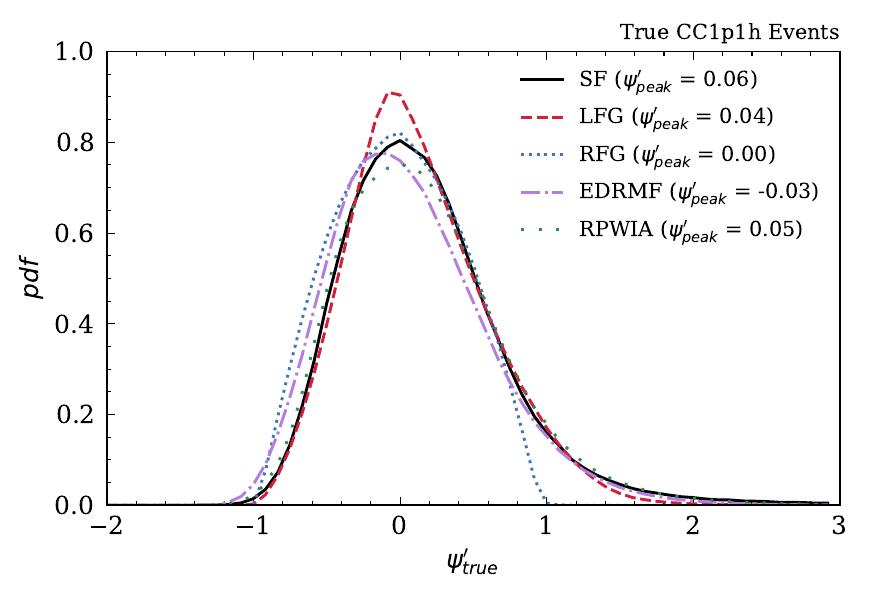}
\caption{The true superscaling variable for true CC1p1h events, for the five models. The legend shows the peak positions for each model.}
\label{Fig:PsiT_all}
\end{center}
\end{figure}

Figure \ref{Fig:PsiT_all} shows the distribution of the superscaling variable from the event kinematics of the five models, where $\psi'$ is calculated using the true $\omega$ and $\vec{q}$ and with a fixed $E_{\rm shift}$ of 20~MeV. The estimated peak position of $\psi'$ for each model is obtained by taking the mean of a Gaussian fit to the core of the $\psi'$ distribution (maximum-1$\sigma$, maximum+1$\sigma$). 

Since the definition of $\psi'$ (Eq.~\ref{eq:psi}) is derived from the RFG model, the distribution of $\psi'$ in the RFG follows the expected behaviour with a peak at zero contained between $\pm1$. The deviation from the assumptions of RFG, particularly when it comes to the momentum of the target nucleon and removal energy distribution, creates the positive tails observed in the other models.

\begin{figure}[h]
\begin{center}
\includegraphics[width=0.45\textwidth]{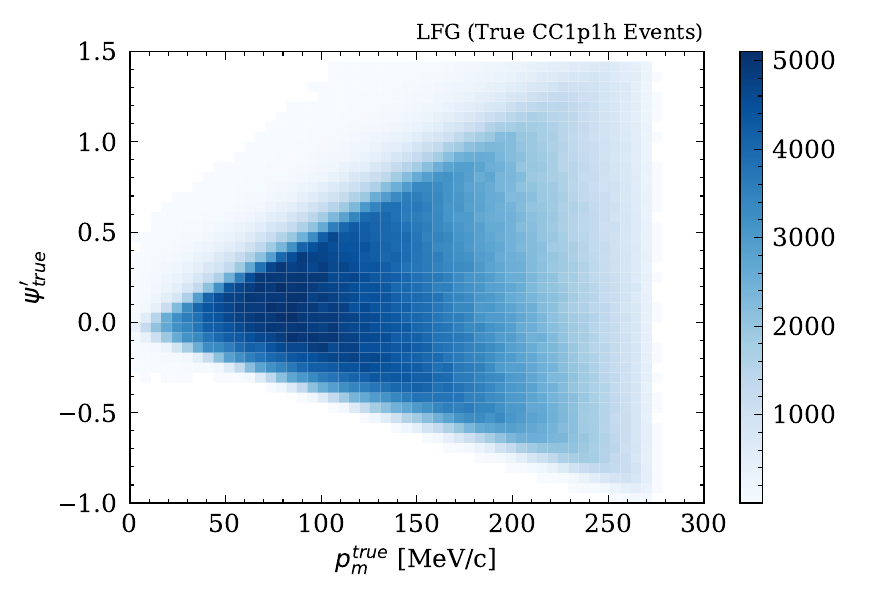}
\caption{Distribution of events as a function of the superscaling variable and the missing momentum. The width of $\psi'$ increases for larger values of $p_m$.}
\label{Fig:PsiTvsPnT}
\end{center}
\end{figure}

\begin{figure}[h]
\begin{center}
\includegraphics[width=0.45\textwidth]{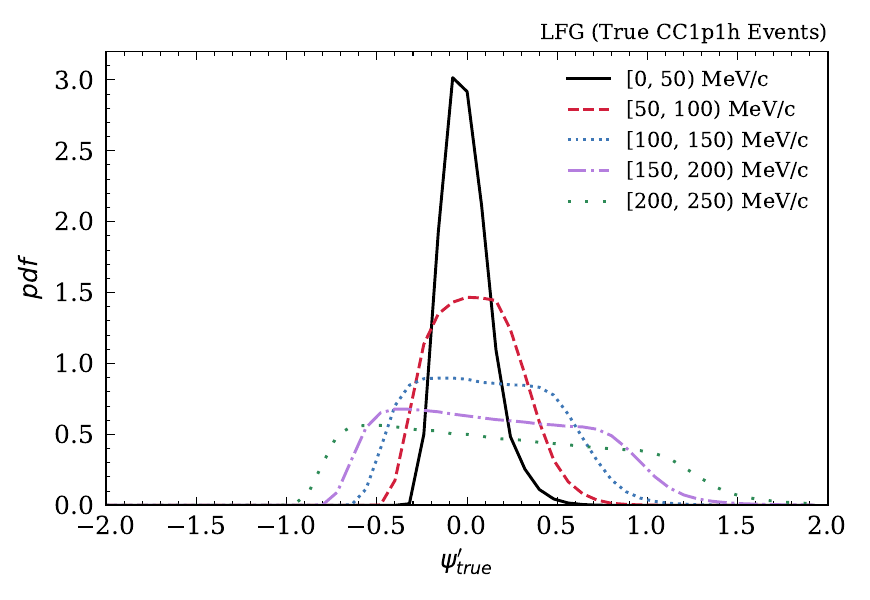}
\caption{Distribution of the superscaling variable in different slices of missing momentum.}
\label{Fig:PsiT_nPint}
\end{center}
\end{figure}

Further details can be inferred from the relation between $\psi'$ and the true missing momentum, equivalent to the momentum of the target neutron in the case of LFG, as shown in Fig.~\ref{Fig:PsiTvsPnT}. According to Eq.~\ref{eq:S-Eshift_noFermi}, in the absence of Fermi momentum ($p_m=0$~MeV/c) the condition for $\psi'=0$ entails that the difference $S - E_{\rm shift}$ can be up to 1.29~MeV. When using the true $\psi'$, we do not make any assumptions on the average removal energy, but we fix $E_{\rm shift}$ to 20~MeV. This gives a narrow $\psi'$ distribution around zero, as seen in Fig.~\ref{Fig:PsiT_nPint}, with a width that is correlated to the width of the removal energy distribution (Fig.~\ref{Fig:TFTN}, bottom plot). The small displacement of the peak position from zero for low $p_m$ can be attributed to the difference $S - E_{\rm shift}$ being larger than 1.29~MeV when using $E_{\rm shift}=20$~MeV while the average removal energy for LFG is 28.39~MeV.

When taking slices of higher $p_m$, as shown in Fig.~\ref{Fig:PsiT_nPint}, 
the distribution of $\psi'$ flattens and does not exhibit a clear peak anymore. This occurs because in the LFG the separation energy is not a single fixed number for each $p_m$, and its dependence on $p_m$ is different than what is found in the RFG.

\begin{figure}[h]
\begin{center}
\includegraphics[width=0.45\textwidth]{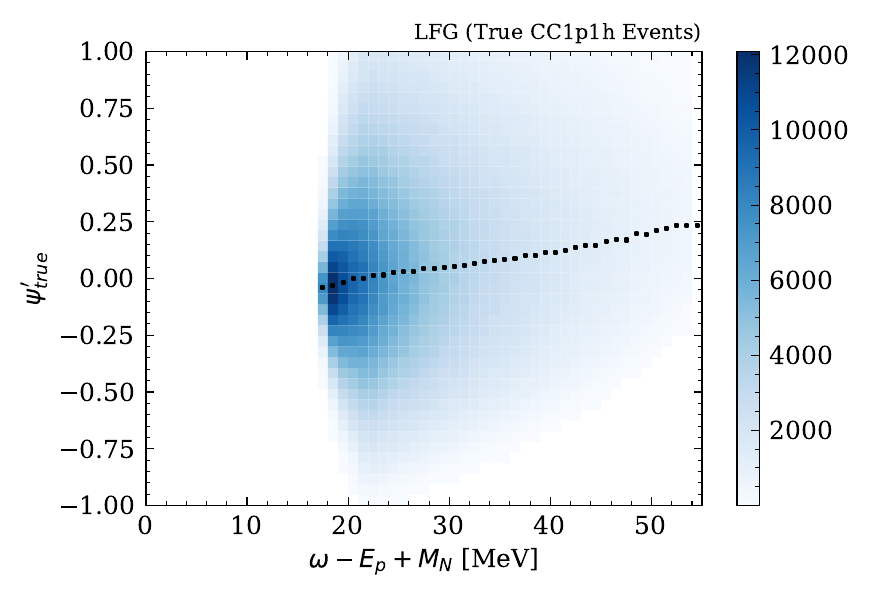}
\caption{The true superscaling variable as a function of the removal energy defined as $\omega - E_p + M_N$ for the LFG model. The points represent the peak position of $\psi'$ for each bin of the removal energy.}
\label{Fig:PsiTvsST}
\end{center}
\end{figure}

\begin{figure}[h]
\begin{center}
\includegraphics[width=0.45\textwidth]{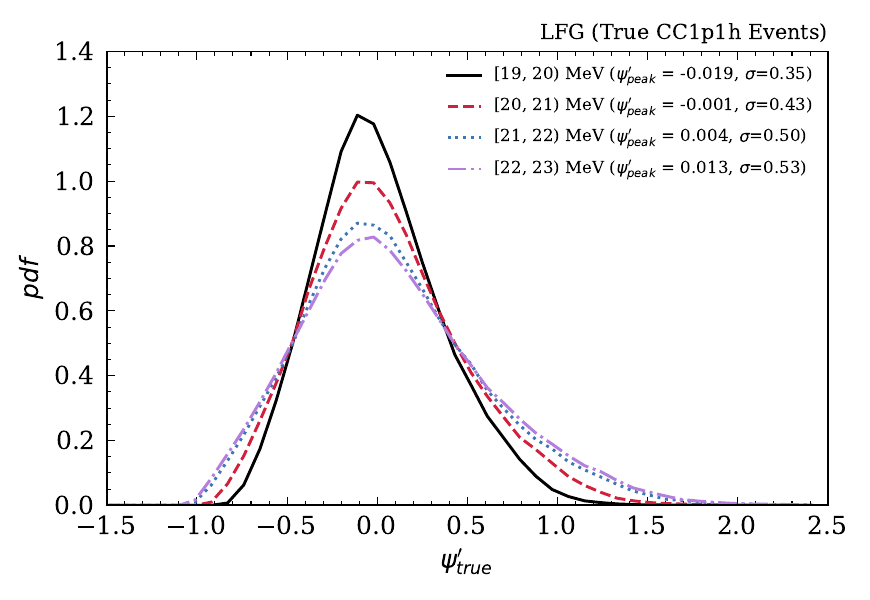}
\caption{The true $\psi'$ distribution for different slices of the removal energy for the LFG model.}
\label{Fig:PsiT_Sint}
\end{center}
\end{figure}

Figure \ref{Fig:PsiTvsST} shows the event distribution of the LFG model in terms of the true $\psi'$ and the removal energy defined in Eq.~\ref{eq:removalEn} and calculated using true event kinematics. The plot shows that $\psi'$ goes to zero when the removal energy is around 20~MeV (equal to $E_{\rm shift}$). For large values of $\omega - E_p + M_N$, there is a bias towards high values of $\psi'$ which produces an asymmetric distribution, as seen in Fig.~\ref{Fig:PsiT_all}, and shifts the peak position towards larger values with a maximum shift of 0.25. The shift in peak position, as well as the increasing asymmetry of $\psi'$ for increasing slices of removal energies can seen in Fig.~\ref{Fig:PsiT_Sint}. For events in these regions, a larger $E_{\rm shift}$ than the one used would be required to shift the QE peak to $\psi' = 0$.

The widening of the $\psi'$ distribution with increasing removal energy is correlated to the missing momentum through the model-dependent relation between $S$ and $p_m$. In the SF model, all values of $p_m$ are accessible in each slice of $S$, and therefore this widening effect is not present. 

\begin{figure}[h]
\begin{center}
\includegraphics[width=0.45\textwidth]{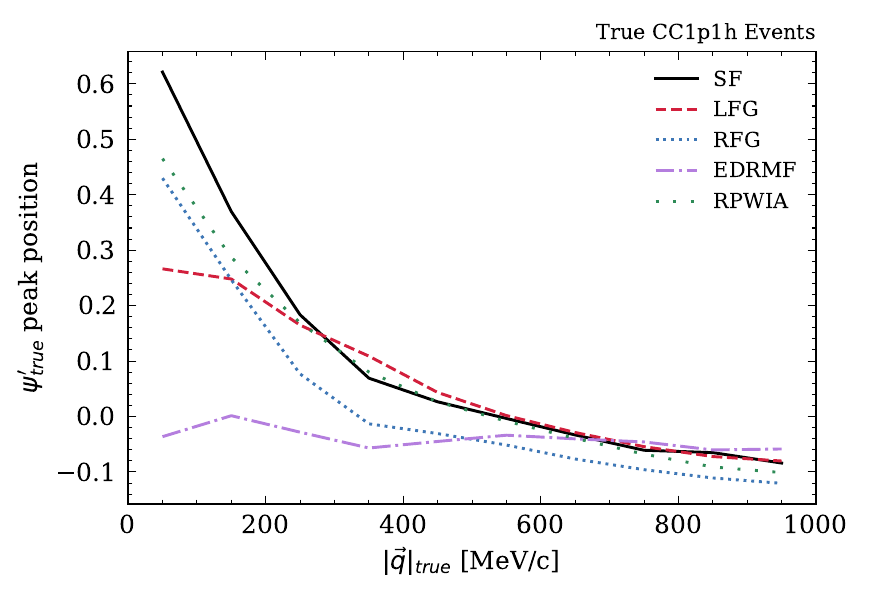}
\caption{The peak positions of the true $\psi'$ distribution as a function of 100~MeV/c slices of the true momentum transfer $|\vec{q}|$ for the five models.}
\label{Fig:PsiT_vc_q3}
\end{center}
\end{figure}

Another source of difference between the models is the relation between the distribution of $\psi'$ and the momentum transfer, $\vec{q}$. In Fig.~\ref{Fig:PsiT_vc_q3} we take 100~MeV/c slices of the true $|\vec{q}|$ and plot the $\psi'$ peak position as a function of the mid point for each slice. In the SF, LFG, RFG and RPWIA models, the peak position of the distribution of $\psi'$ is seen to have a similar dependence on $|\vec{q}|$, namely a positive shift in the peak position at low-$|\vec{q}|$. The EDRMF model is the outlier here, the peak position shows little dependence on the momentum transfer. 
This is to be attributed to the final-state potential included in this calculation. Indeed, a similar difference in the peak position of the scaling function in the RPWIA and EDRMF approaches as function of $|\vec{q}|$ can be seen for electron-scattering cross sections in Ref.~\cite{PhysRevC.101.015503}.

\begin{figure}[h]
\begin{center}
\includegraphics[width=0.45\textwidth]{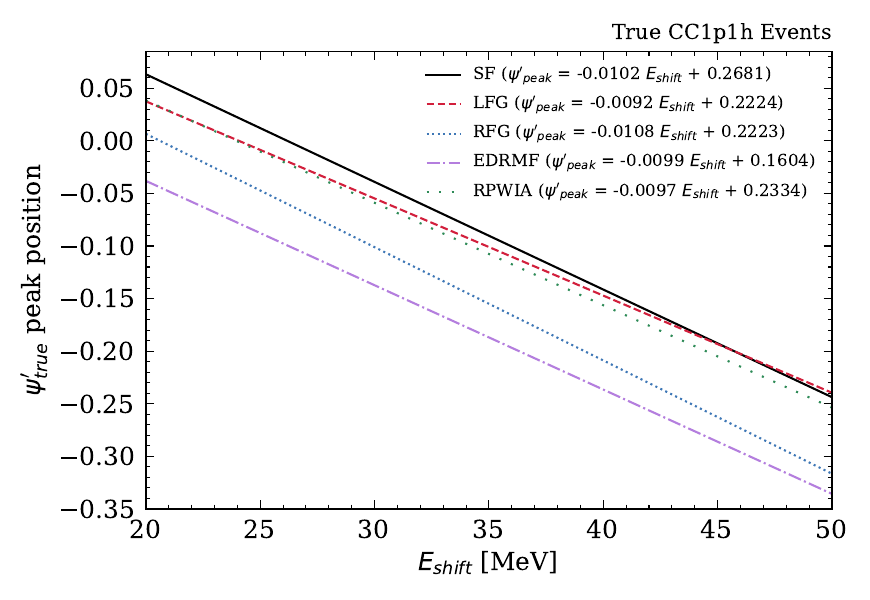}
\caption{The peak positions of each $\psi'$ distribution as a function of the $E_{\rm shift}$ value used in its calculation for the five models.}
\label{Fig:PsiT_EshiftScan}
\end{center}
\end{figure}

Finally, we performed a search for the $E_{\rm shift}$ value that gives a $\psi'$ distribution peaking at zero for $E_{\rm shift}$ between 20 and 50 MeV. Figure~\ref{Fig:PsiT_EshiftScan} shows that the relation between the peak position and $E_{\rm shift}$ is approximately linear, with a nearly universal slope for the models, but differing intercepts. 

The $E_{\rm shift}$ values that give a $\psi'$ distribution peaking at zero for each model are listed in Tab.~\ref{tab:psiT_EshiftScan}. As expected, the RFG model arrives at $\psi'=0$ at an $E_{\rm shift}$ value closest to the 20~MeV used in the calculation~\cite{PsiParameters}, while the other models have different results based on their individual removal energy distributions. Comparing these values with the average removal energy (the means in Fig.~\ref{Fig:TFTN}), we observe that they are within a few MeVs, except for the RFG model where $\overline{\omega - E_p + E_N}$ is more consistent with the model's assumptions, and the EDRMF model where the final-state potential is known to shift the position of the quasi-elastic peak to smaller energy transfer, in line with electron scattering data~\cite{Gonz_lez_Jim_nez_2019}. This means that the distribution peaks at lower $\psi'$, as seen in Fig.~\ref{Fig:PsiT_all}. 

\begin{table}[h]
    \centering
    \caption{The values of $E_{\rm shift}$ that give a $\psi'_{\rm true}$ distribution peaking at zero for each model, compared to the average removal energy defined as $\overline{\omega - E_p + M_N}$.}
    \begin{tabular}{l|c|c}
    \hline \hline
    \rule{0pt}{2.2ex}\textbf{Model} & \textbf{$E_{\rm shift}|_{\psi'=0}$} [MeV]  & \textbf{$\overline{\omega - E_p + M_N}$} [MeV]\\[1ex] \hline
        SF & 26.19 & 26.45\\
        LFG & 24.07 & 28.39\\
        RFG & 20.62 & 9.91\\
        EDRMF & 16.16 & 29.02\\
        RPWIA & 23.95 & 29.20\\
    \hline \hline
    \end{tabular}
    \label{tab:psiT_EshiftScan}
\end{table}

This measurement of $E_{\rm shift}$ relies on using true Monte Carlo (MC) information to obtain the true $\psi'$ distribution. This is not achievable experimentally since energy and momentum transfer are not observable in a neutrino experiment. However, by taking a fixed average value of the removal energy ($S$), we can reconstruct the full kinematics of a 1$\mu$1p event, and obtain an approximate superscaling variable. We will later perform another measurement of $E_{\rm shift}$ using the reconstructed $\psi'$ and under more experimental-like conditions.

\section{ Experimental-like Event Selection } 
\label{sec:expEvSel}
For a more realistic sample of events from an experimental point of view, we use the selection criteria for the CC0$\pi1p$ topology as done for the T2K experiment~\cite{T2K:2018rnz}. This selection requires the detection of a muon, and a single proton in the final state, with $|\vec{k_{p}}| > 0.45$ GeV and $\cos{\theta_{p}} > 0.4$ to reflect the proton detection threshold and detector acceptance. Events with multiple protons, or pions, exiting the nucleus are rejected.

This selection is applied to events following nuclear re-scattering, where we use the intra-nuclear cascade model implemented in NEUT for all five models.

For the NEUT models (SF, LFG and RFG), we distinguish five different samples that pass our selection according to their production mechanism: 
\begin{enumerate}
    \item CC1p1h, charged current one particle one hole, when the proton emerges from the nucleus with no NrS. 
    \item CC1p1h+NrS, charged current one particle one hole, when the proton undergoes NrS. 
     \item CC2p2h, charged current two particles two holes where the second proton is undetectable.
     \item CCRes, charged current resonance, where the pion is absorbed by the nucleus.
     \item CCDis, charged current deep inelastic scattering, where all pions are absorbed by the nucleus.
\end{enumerate}

In the case of EDRMF and RPWIA, only CC1p1h and CC1p1h+NrS are included as they are not native NEUT models.

The modeling of the selection and the prediction from the LFG model in NEUT are in very good agreement with T2K and MINERvA experimental results as shown in \cite{Bourguille:2020bvw}.

\begin{figure}[h]
\begin{center}
\includegraphics[width=0.45\textwidth]{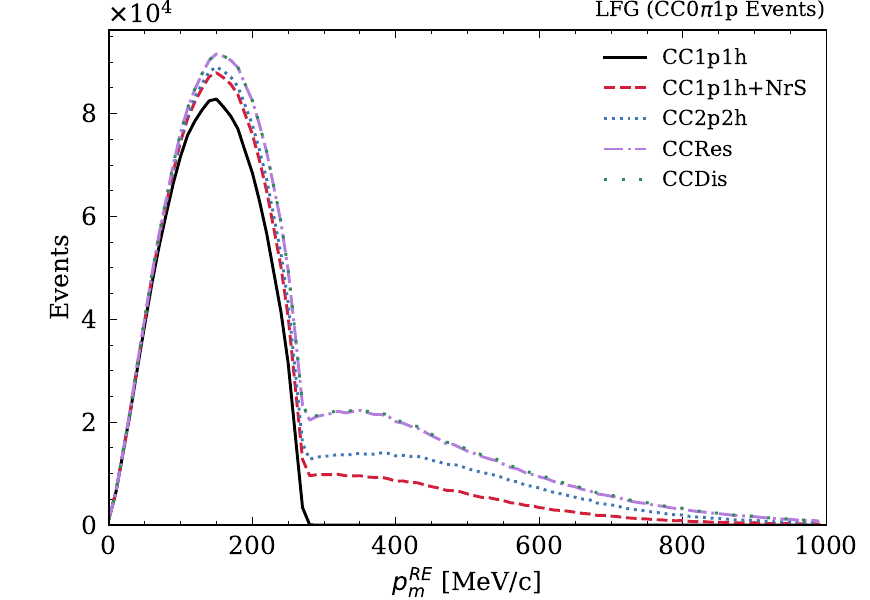}
\caption{A stacked histogram of the reconstructed missing momentum via RE for the LFG model.}
\label{Fig:Pnreco}
\end{center}
\end{figure}

Additionally, we also introduce a cut on the reconstructed missing momentum as proposed in \cite{Furmanski:2016wqo} and the first experimental applications described in \cite{MicroBooNE:2023wzy}. When including background events that pass the CC0$\pi1p$ selection, the reconstructed missing momentum shows a very distinctive shape with CC1p1h events concentrated below the Fermi level (at around 300~MeV/c), see Fig.~\ref{Fig:Pnreco}. This variable, thus, provides a very efficient cut to reduce the contamination of NrS, CC2p2h and CCRes and increase the purity of CC1p1h events in the selected sample. 

Table ~\ref{tab:pn_cut} shows the detailed sample composition for the five models before and after introducing the cut at $p_m^{\rm RE}=300$~MeV/c.

\begin{table}[h]
    \caption{The fraction of each interaction type in the original sample of events (top row) compared to the sample after applying the cut on the missing momentum (bottom row) for each model.}
    \centering
    \resizebox{\linewidth}{!}{%
    \begin{tabular}{l|c|c|c|c|c}
    \hline \hline
    \textbf{Model} & \textbf{CC1p1h} & \textbf{CC1p1h+NrS} & \textbf{CC2p2h} & \textbf{CCRes} & \textbf{CCDis}\\ \hline
        \multirow{2}{*}{SF} & 0.575 & 0.143 & 0.148 & 0.127 & 0.007\\
           & 0.833 & 0.076 & 0.032 & 0.057 & 0.001\\
        \hline
        \multirow{2}{*}{LFG} & 0.560 & 0.171 & 0.097 & 0.125 & 0.007\\
            & 0.839 & 0.085 & 0.023 & 0.052 & 0.001\\
        \hline
        \multirow{2}{*}{RFG} & 0.636 & 0.164 & 0.085 & 0.109 & 0.006\\
            & 0.860 & 0.075 & 0.019 & 0.044 & 0.001\\
        \hline
        \multirow{2}{*}{EDRMF} & 0.764 & 0.236 & -- & -- & --\\
            & 0.897 & 0.103 & -- & -- & --\\
        \hline
        \multirow{2}{*}{RPWIA} & 0.770 & 0.230 & -- & -- & --\\
            & 0.901 & 0.099 & -- & -- & --\\
       
    \hline \hline
    \end{tabular}}
    \label{tab:pn_cut}
\end{table}

\section{Experimental Observables and the Reconstructed Superscaling Variable}
In this section we take a practical point of view, and discuss a measurement of the reconstructed superscaling variable, based on the energy reconstruction in 1$\mu$1p events presented in section~\ref{sec:recon}.
We make use only of observables accessible in neutrino detectors, and include re-scattering in the intra-nuclear cascade model as discussed in the previous section.

\subsection{The Superscaling Variable}
\begin{figure}[h]
\begin{center}
\includegraphics[width=0.45\textwidth]{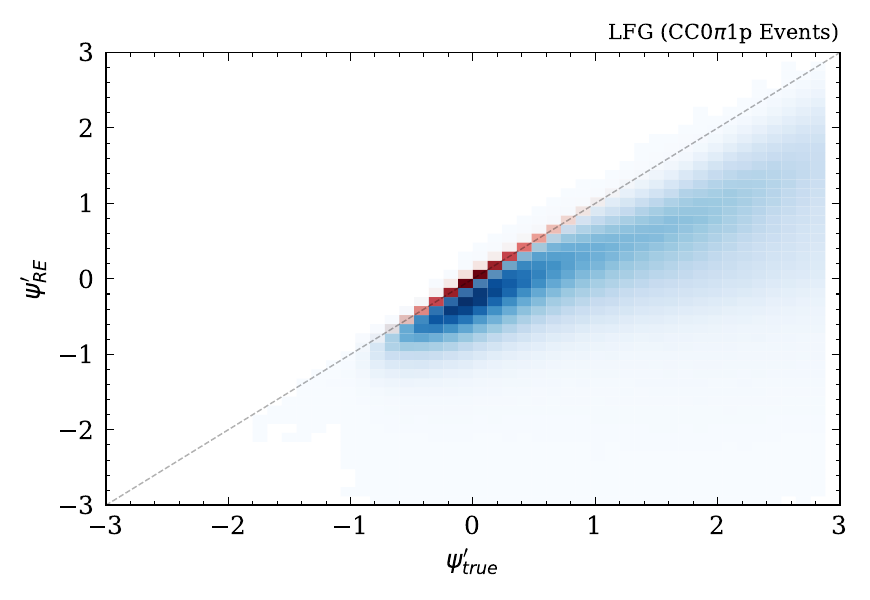}
\caption{Reconstructed vs true $\psi'$ for the LFG model, where event kinematics were reconstructed via RE and no cuts were applied to minimize the background. True CC1p1h events (red) fall on the diagonal, while all other events (blue) have lower reconstructed $\psi'$ compared to the true value.}
\label{Fig:PsivsPsi}
\end{center}
\end{figure}

\begin{figure}[h]
\begin{center}
\includegraphics[width=0.45\textwidth]{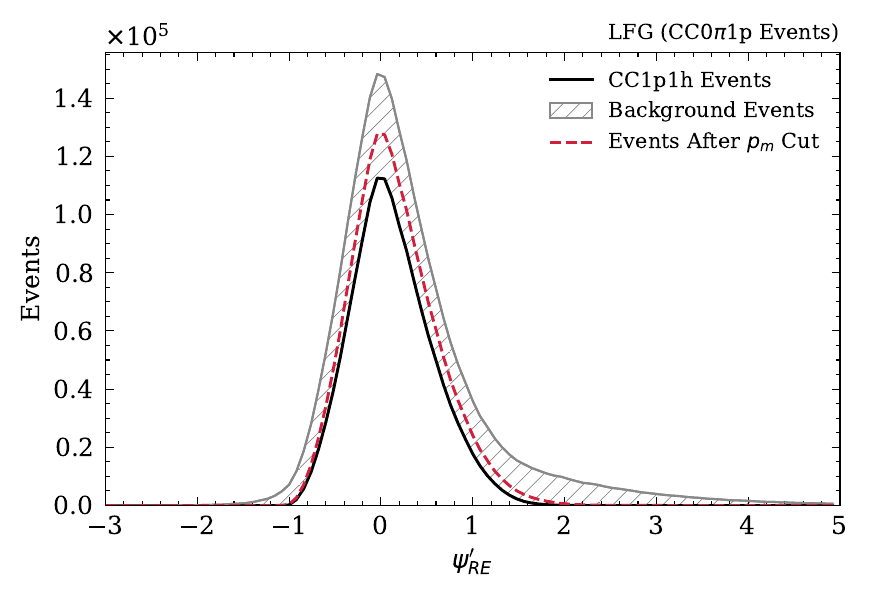}
\caption{The reconstructed superscaling variable for the LFG model. Applying the cut $p_{m}^{\rm RE} < 300$ MeV/c eliminates a good fraction of the background reducing the positive tails.}
\label{Fig:PsiRE_Pncut}
\end{center}
\end{figure}

The reconstructed superscaling variable ($\psi'_{\rm RE}$) can be calculated from Eq.~\ref{eq:psi} following the RE energy reconstruction to obtain the energy and momentum transfer. Figure \ref{Fig:PsivsPsi} shows the $\psi'_{\rm RE}$ distribution for the LFG model as a function of the true value for all events passing the CC$0\pi1p$ selection. We observe strong correlation between the true and reconstructed values corresponding to CC1p1h events (distinguished in red). Under the CC1p1h line, the correlation disappears for background events (in blue) as the reconstruction does not account for their energy losses resulting in the wrong energy-momentum transfer estimation and leading to an underestimated $\psi'_{\rm RE}$ calculation. This can shift the peak position of $\psi'_{\rm RE}$ and produce negative tails. However, imposing the cut on the reconstructed missing momentum can be used to limit these events. The distribution of $\psi'_{\rm RE}$ before and after the cut for the LFG model is shown in Fig.~\ref{Fig:PsiRE_Pncut}, where the purity of CC1p1h (no NrS) events has increased from 0.56 to 0.84. We will apply this cut to the selected sample of events in what follows.

\begin{figure}[h]
\begin{center}
\includegraphics[width=0.45\textwidth]{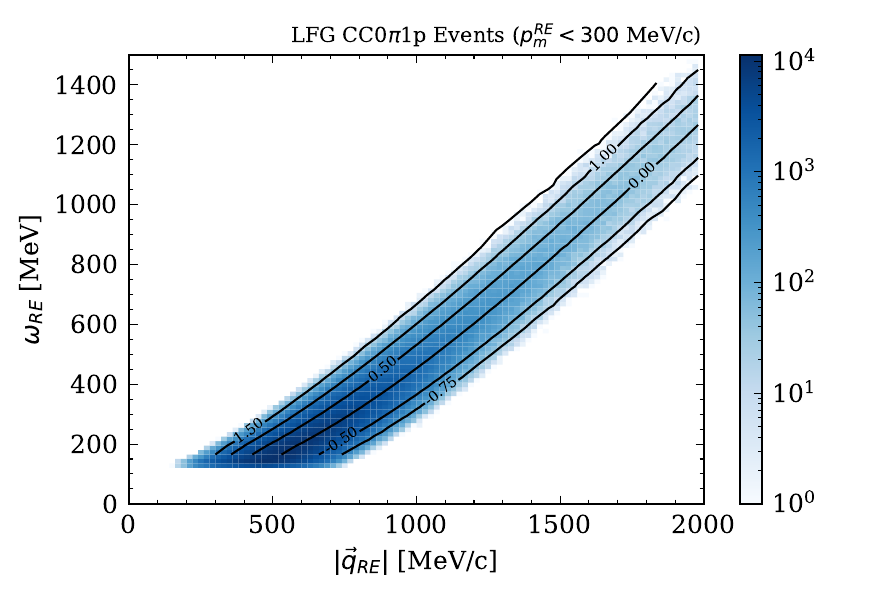}
\caption{ The reconstructed energy transfer $\omega$ versus the reconstructed momentum transfer $|\vec{q}|$. Overlaid are the contour lines representing iso-$\psi'_{\rm RE}$ values.}
\label{Fig:Q0Q3}
\end{center}
\end{figure}

 \begin{figure}[h]
\begin{center}
\includegraphics[width=0.45\textwidth]{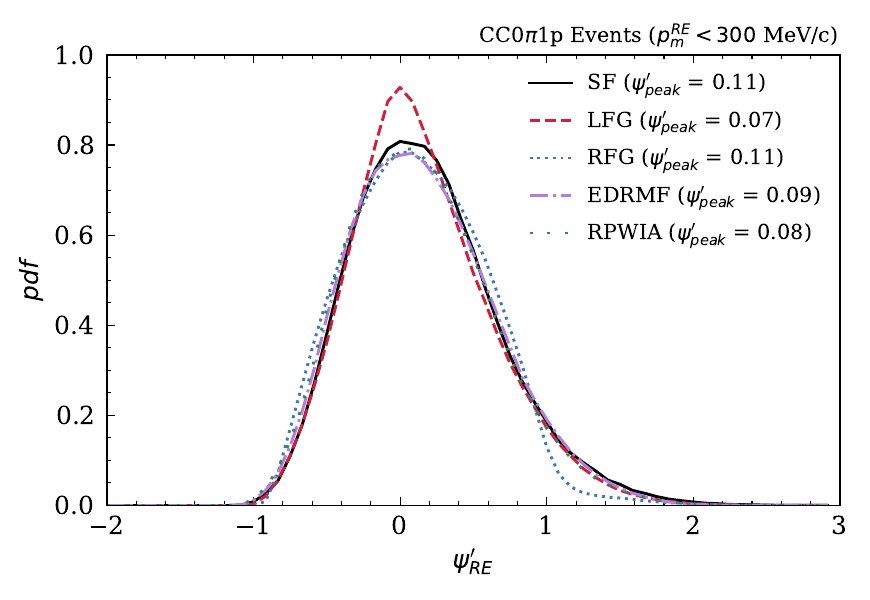}
\caption{The reconstructed superscaling variable for the five models. The cut on $p_m^{\rm RE}$ was applied to minimize the background.}
\label{Fig:PsiAllModels}
\end{center}
\end{figure}

The values of the reconstructed energy and momentum transfers used in the calculation of $\psi'_{\rm RE}$ are shown in Fig.~\ref{Fig:Q0Q3}. The cutoff at low $\omega$ is a consequence of the requirement of a 450~MeV/c proton in the final state, which corresponds to a kinetic energy of about 140~MeV. The non-linearity of the iso-$\psi'_{\rm RE}$ lines means that negative values of the superscaling variable are more affected by this limitation on the proton momentum than positive ones, which introduces another source of asymmetric in the distribution of $\psi'_{\rm RE}$. Another consequence of the acceptance cut on the proton momentum is that it eliminates a region of phase space that particularly highlights differences between the models in their $\vec{q}$ dependence (low $|\vec{q}|$ in Fig.~\ref{Fig:PsiT_vc_q3}) as well as in their treatment of Pauli blocking. This is a delicate experimental issue that will be partially circumvented by the new generation of near detectors~\cite{Dolan:2021hbw} that reduce the proton detectability threshold. 

Using the RE method, and the aforementioned selection and $p_m^{\rm RE}$ cut on all models, Fig.~\ref{Fig:PsiAllModels} shows the $\psi'_{\rm RE}$ distributions for the five considered models. Compared to the true distributions and peak positions (Fig.~\ref{Fig:PsiT_all}), all reconstructed distributions exhibit the positive shifts discussed above. The average shift in the peak position for all five models is 0.068, with maximum shifts observed in the EDRMF and RFG models. 

In the case of the RFG, the large difference between true and reconstructed $\psi'$ is mainly due to the energy reconstruction. On the other hand, the final state potential of the EDRMF model—previously shown to cause $\psi'$ to peak at lower values in section~\ref{sec:truePred}—has a diminishing effect when the proton detection threshold is applied and the low $|\vec{q}|$ region is eliminated. This causes the peak position of the EDRMF model to align with that of the RPWIA.

\begin{figure}[h]
\begin{center}
\includegraphics[width=0.45\textwidth]{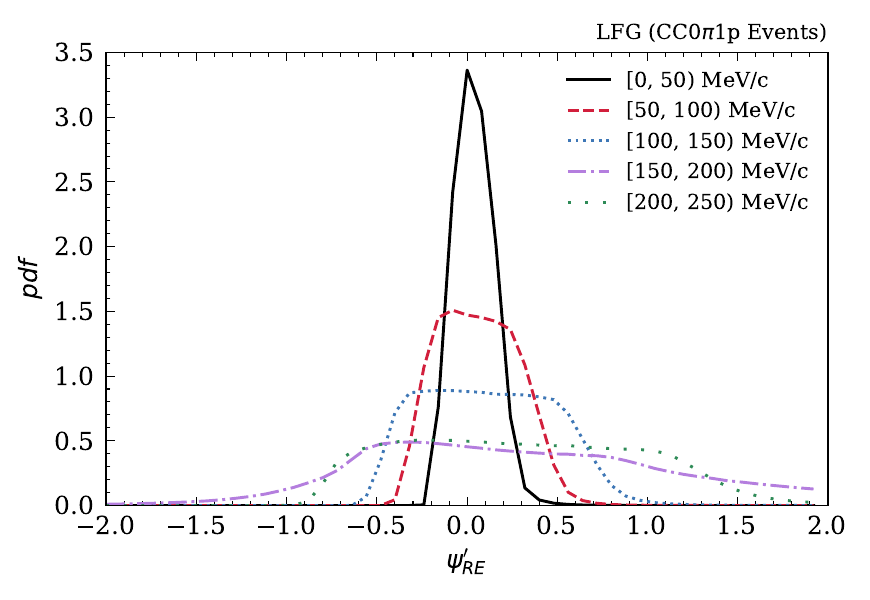}
\caption{The reconstructed superscaling variable for different slices of the reconstructed missing momentum. $\psi'_{\rm RE}$ becomes more flat for increasing $p_m^{\rm RE}$.}
\label{Fig:PsiRE_nPint}
\end{center}
\end{figure}

In the true Monte Carlo predictions we observed that the distribution of $\psi'$ loses its distinguished peak and becomes more flat when taking slices of increasing missing momenta. This effect can also be observed in the reconstructed case, see Fig.~\ref{Fig:PsiRE_nPint}, and would affect the shape of the $\psi'_{\rm RE}$ distribution for experimental conditions where only a certain part of the kinematical phase space is accessible. A good reconstruction of the missing momentum allows for identifying the regions with low $p_m$ where the $\psi'$ distribution is more narrow and the peak position can be determined with less bias.

\subsection{The Removal and Shift Energies}
\begin{figure}[h]
\begin{center}
\includegraphics[width=0.45\textwidth]{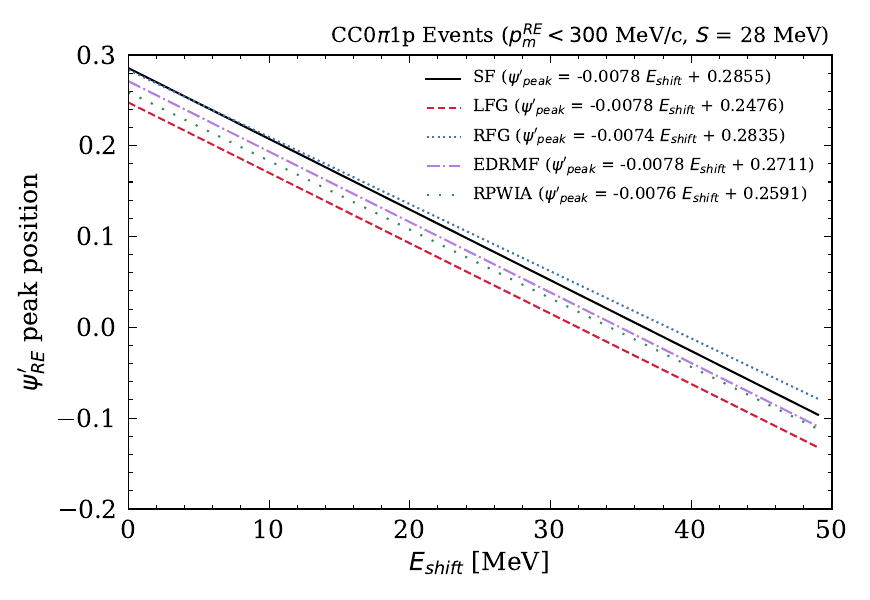}
\caption{The peak position of the reconstructed $\psi'$ as a function of $E_{\rm shift}$. The average removal energy is fixed to 28~MeV in the energy reconstruction for all models.}
\label{Fig:PsivsS_reco_AllModels}
\end{center}
\end{figure}

Using $\psi'_{\rm RE}$, we can perform another search for the $E_{\rm shift}$ values required to shift the peak position to zero for each model. Figure \ref{Fig:PsivsS_reco_AllModels} shows the peak positions of $\psi'_{\rm RE}$ for each value of $E_{\rm shift}$. Compared to Fig.~\ref{Fig:PsiT_EshiftScan}, the slopes for all five models are now closer in value. This is a consequence of the detectability threshold of the proton momentum, which eliminates the low $|\vec{q}|$ region, minimizing the differences between models and leading to more similar slope values.

To distinguish between the effect of the energy reconstruction and that of the sample selection and detector acceptance cuts, we perform the same exercise using the reconstructed energy and momentum transfer but on true CC1p1h events. Table~\ref{tab:psiRE_EshiftScan} shows the $E_{\rm shift}$ values required to shift the $\psi'_{\rm RE}$ peak to coincide with zero for the experimental-like event sample (left) compared to the true CC1p1h event sample (center). On the right, we also include the results obtained in Tab.~\ref{tab:psiT_EshiftScan} for the true $\psi'$.

\begin{table}[htp]
    \caption{The values of $E_{\rm shift}$ that give a $\psi'_{\rm RE}$ peaking at zero for a realistic sample of CC0$\pi$1p events (with the $p_m^{\rm RE}$ and proton acceptance cuts) and true CC1p1h events. Compared to the $E_{\rm shift}$ that give a $\psi'_{\rm true}$ peaking at zero.}
    \centering
    \resizebox{\linewidth}{!}{%
    \begin{tabular}{l|c|c|c}
    \hline \hline
    \rule{0pt}{2.2ex}\multirow{2}{*}{\textbf{Model}} & \multicolumn{2}{c|}{\textbf{$E_{\rm shift}|_{\psi'_{\rm RE}=0}$} [MeV]} & {\textbf{$E_{\rm shift}|_{\psi'_{\rm true}=0}$} [MeV]}\\ [1ex]
    \cline{2-4}
   \rule{0pt}{2.3ex}  & CC0$\pi$1p* & CC1p1h & CC1p1h\\
     \hline
        SF & 36.64 & 25.28 & 26.19\\
        LFG & 31.94 & 25.59 & 24.07\\
        RFG & 38.34 & 25.37 & 20.62\\
        EDRMF & 34.93 & 16.24 & 16.16\\
        RPWIA & 34.23 & 21.88 & 23.95\\
    \hline \hline
    \multicolumn{4}{l}{\footnotesize *CC0$\pi$1p with proton acceptance cuts and $p_m^{\rm RE}<$300~MeV/c.}
    \end{tabular}}
    \label{tab:psiRE_EshiftScan}
\end{table}

We observe that larger $E_{\rm shift}$ values are required for $\psi'_{\rm RE}$ in the experimental-like event sample. This is to compensate for the effect observed in Fig.~\ref{Fig:Q0Q3} where the proton acceptance cut eliminates more events on the negative side of $\psi'$ compared to the positive, as well as the inclusion of NrS and CC0$\pi$1p background events that occupy the positive side of $\psi'$. 

On the other hand, for true CC1p1h events without any cuts, we obtain similar values of $E_{\rm shift}$ for the true and reconstructed $\psi'$. This means that the energy reconstruction and our assumption on the average removal energy has a small effect on the $\psi'_{\rm RE}$ peak position and the required $E_{\rm shift}$, and the error is dominated by the proton's momentum detection threshold and background contamination.

\section{ Conclusions } 
We presented the superscaling variable as an observable in charged current neutrino-nucleus interactions, comparing the shape and characteristics of the $\psi'$ distribution for five neutrino-nucleus interaction models. Using true Monte Carlo information, we studied the superscaling variable dependencies with fundamental parameters of the models such as the removal energy, the target nucleon momentum and the momentum transfer of the interaction.

We have shown that $\psi'$, and particularly the peak position of the distribution, provides information on the removal energy of nucleons. In the peak region, where $\psi'=0$, the average removal energy is related to the shift energy, $E_{\rm shift}$, required to shift the quasi-elastic peak to zero, at low missing momentum

To comment on the determination of $\psi'$ in neutrino interaction experiments where the neutrino energy is not directly measurable, we introduced two neutrino energy reconstruction methods built on different assumptions for the average removal energy. We concluded that reconstruction by energy-momentum conservation (REP), where the missing energy and momentum are fully correlated, performs better for the RFG model due to its simplicity, while reconstruction by energy conservation (RE), where the missing (removal) energy is assumed to be constant, has the advantage in all other models.

Applying the RE method on a more realistic event sample that passes the CC0$\pi$1p selection, and utilizing a cut on the missing momentum to increase the purity of CC1p1h events, we were able to obtain the reconstructed superscaling variable with an average displacement of 0.068 for the peak position with respect to the true distributions among the five models.

Finally, we have shown that while the energy reconstruction is sufficient to measure $\psi'$ experimentally, the limitations of current detector technologies in measuring low momentum protons impose a bias on the $\psi'$ distribution requiring a larger $E_{\rm shift}$ to make the peak occur at $\psi'=0$. Even though it is difficult to estimate the true average removal energy, the procedure outlined here—due to the simplicity of the energy reconstruction method and robustness of the results—allows to make a novel model-independent measurement which may be used to constrain or distinguish interaction models. Additionally, future and upgraded detectors are expected to have lower detection thresholds that allow for more accurate measurements~\cite{Dolan:2021hbw}.

\begin{acknowledgments}
This work was supported by the Swiss National Foundation Grant No. 200021\_85012 and Fermi Research Alliance, LLC under Contract No. DE-AC02-07CH11359 with the U.S. Department of Energy.
\end{acknowledgments}

\newpage
\bibliography{biblio}
\end{document}